\documentclass[prl,amsmath,superscriptaddress,amssymb,twocolumn,floatfix,aps,showpacs,showkeys]{revtex4-1}
\usepackage{amssymb}   
\usepackage{amssymb}
\usepackage{epsfig}
\usepackage{graphicx} 
\usepackage{multirow}
\usepackage[space]{grffile}
\usepackage{color}

\begin{document}
\title{Physical properties of resonances as the building blocks\\ of multichannel scattering amplitudes}

\author{S.~Ceci}
\email{sasa.ceci@irb.hr}
\affiliation{Rudjer Bo\v{s}kovi\'{c} Institute, Bijeni\v{c}ka  54, HR-10000 Zagreb, Croatia}
\author{M.~Vuk\v si\' c}
\affiliation{University of Zagreb, Bijeni\v{c}ka  34, HR-10000 Zagreb, Croatia}
\author{B.~Zauner}
\affiliation{Rudjer Bo\v{s}kovi\'{c} Institute, Bijeni\v{c}ka  54, HR-10000 Zagreb, Croatia}


\begin{abstract}
We propose a simple formula for multichannel resonant scattering with parameters related to physical resonant properties. It can be used to predict residue phase from other resonant parameters and describe the shape of scattering amplitudes close to the resonance without any background contributions or fitting. It works well even for overlapping resonances. Imposing unitarity to the proposed formula enabled us to explain some puzzling features of much more advanced models. 
\end{abstract}

\keywords{Resonance mass, Scattering amplitude poles, Breit-Wigner parameters}
\pacs{11.55.Bq, 14.20.Gk, 11.80.Gw, 13.75.Gx}

\maketitle

\section{Introduction}

In 1936, Gregory Breit and Eugene Wigner wrote a paper on the capture of slow neutrons in nuclei \cite{BW}. There they produced a famous equation for the resonant scattering, known now as the Breit-Wigner formula. It can be found in most textbooks on molecular physics \cite{AtkinsFriedman}, nuclear physics \cite{NuclPh}, and quantum field theory \cite{QFT}. It is also present in {\em The Review of Particle Physics} \cite{PDG}, a publication issued every two years by the {\em Particle Data Group} (PDG). 

The formula is often expressed through scattering amplitude $A$ as a function of energy $W$ 
\begin{equation}\label{OriginalBW}
A= \frac{x\, \Gamma/2}{M-W-i\,\Gamma/2},
\end{equation}
where $M$ is the resonant mass, $\Gamma$ the total decay width, and $x$ is the branching fraction. 

For the elastic resonances, those decaying only to the same particles they are produced from, $x$ equals one and the formula is manifestly unitary. By unitary, we mean that the scattering matrix that is built up from the scattering amplitude as \mbox{$1+2\,iA$} will be a unitary matrix. 

The Breit-Wigner formula has been substantially modified since then to better represent the realistic scattering amplitudes with their angular-momentum-dependent threshold behaviors, various background contributions, and multiple, sometimes even strongly overlapping resonances. Representative and oft-quoted examples of such modifications are in the pion-pion \cite{GounarisSakurai,Flatte} and the pion-nucleon scattering \cite{Cutkosky,ManleySaleski,Batinic}. 

In addition to parameters $M$, $\Gamma$, and $x$, these modifications produce another set of resonant properties. They are sometimes called the conventional resonant parameters, but are better known as the Breit-Wigner parameters: mass $M_\mathrm{BW}$, width $\Gamma_\mathrm{BW}$, and branching fraction $x_\mathrm{BW}$.  Parameters $M$ and $\Gamma$ are now known as the pole parameters, with the pole position equal to \mbox{$M-i\,\Gamma/2$}. 

It is important to stress that the Breit-Wigner mass is not just some model parameter in a modified Breit-Wigner formula, as it is explained in detail, e.g., in section III.~of Ref.~\cite{Cutkosky}. In fact, it is directly connected with the renormalized resonant mass of a propagator in quantum-field theory \cite{Sirlin,Man95,Tornqvist,Scherer,Cec09}, which is defined as the real energy at which propagator's denominator becomes purely imaginary. 

The pole, on the other hand, might appear like it is just some mathematical feature of scattering matrix, but it is the defining characteristic of the resonance \cite{DalitzMoorhouse}. In addition to the pole position, researchers extract the residue parameters as well. The residue magnitude $|r|$ is simply given by $x$ times $\Gamma/2$. But there is also the residue phase $\theta$. It has no counterpart in the textbook Breit-Wigner formula. Moreover, its physical meaning is much less clear than the meaning of other parameters. The mass is roughly related to the resonant energy, the decay width is inversely related to the resonance lifetime, and the branching fraction gives the probability of a resonance decaying to a particular set of lighter particles. So what is the physical meaning of the residue phase? 

A step towards the possible answer was made in a broad, yet brief study on resonances \cite{Cec13} where a five-parameter semi-empirical Breit-Wigner-like formula was introduced. The relation between the elastic residue phase, elastic threshold, and the Breit-Wigner mass was later established and tested in a study on nucleon resonances \cite{Cec16}. 

In this paper we propose a multichannel version of this formula with parameters directly related to either channel thresholds or physical resonant properties. We test it on prominent nucleon resonances by predicting the residue phase from other resonant parameters \cite{PDG,Sva14}, show that the shape of the predicted amplitude agrees with the data \cite{SAID}, analyze the robustness of the residue phase estimation on an overlapping resonance system \cite{Cec16}, and explain a puzzling feature of an advanced unitary model \cite{Batinic} where the imaginary part of the amplitude matrix trace had peak positions at the Breit-Wigner masses \cite{Cec08}.

\section{Model}

We begin with the scattering amplitude first suggested in Ref.~\cite{Cec13} 
\begin{equation} \label{NonUnitaryAmp}
A = x  \, \sin(\rho+\delta) \, e^{i(\rho +\beta)},
\end{equation}
where 
\begin{equation} \label{rho}
\tan\rho = \frac{\Gamma/2}{M-W}.
\end{equation}
$M$, $\Gamma$, and $x$ are familiar pole parameters, while two new parameters are phases $\beta$ and $\delta$. Note that if they are both zero, we will get exactly Eq.~(\ref{OriginalBW}). 

In Ref.~\cite{Cec16}, the Breit-Wigner mass is defined as the real energy $W$ at which $\rho + \beta$ becomes 90$^\circ$, i.e.,
\begin{equation}\label{beta}
M_\mathrm{BW} = M-\Gamma/2 \, \tan\beta. 
\end{equation}
The elastic phase $\delta$ is related to elastic threshold $W_0$ as 
\begin{equation} \label{delta}
W_0 = M+\Gamma/2 \, \cot\delta.
\end{equation} 
The elastic residue phase is then $\theta=\beta+\delta$.

\begin{figure}[h!]
\begin{center}
\includegraphics[width=0.47\textwidth]{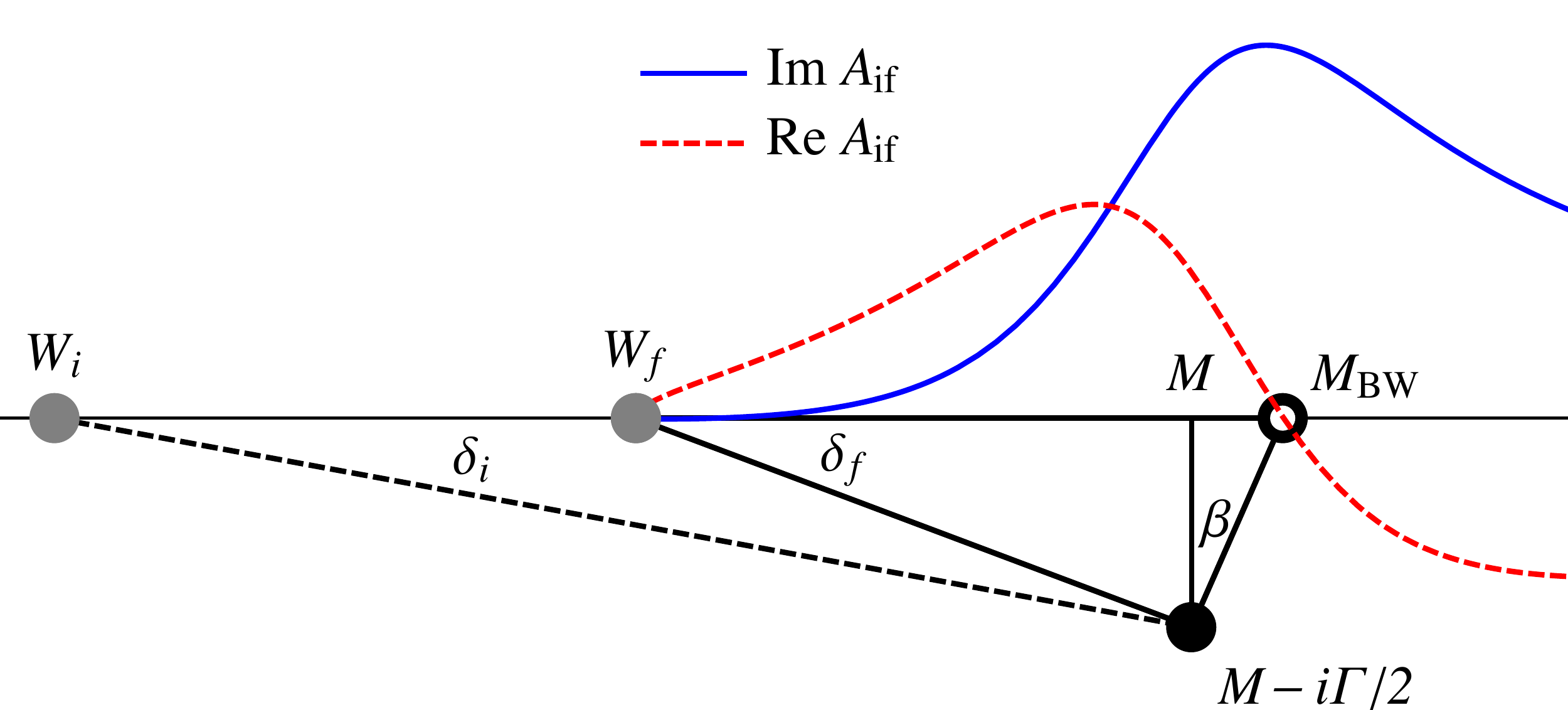}
\caption{Geometrical interpretation of $\beta$ and $\delta$s. All angles are negative here. \label{Fig1}}
\end{center}
\end{figure}

To generalize Eq.~(\ref{NonUnitaryAmp}) for inelastic processes $i\rightarrow f$ we propose a simple modification 
\begin{equation} \label{UnitaryAmp}
A_{if}=x_{if}\, \sqrt{\sin(\rho+\delta_i)\,\sin(\rho+\delta_f)} \, \,e^{i(\rho +\beta)} .
\end{equation}
where $x_{if}=\sqrt{x_i\, x_f}$.  The residue phase of this amplitude is given by 
\begin{equation}\label{UnitaryTheta}
\theta_{if}=\beta+\frac{\delta_i+\delta_f}{2}.
\end{equation}
To estimate $\theta_{if}$, we need $\beta$ and $\delta$s, which are determined from the geometry, as is shown in Fig.~\ref{Fig1}. 

\section{Results}

In Ref.~\cite{Cec16}, residue phases of five prominent nucleon resonances seen in the elastic pion-nucleon scattering are predicted from other resonance parameters using elastic Eq.~(\ref{NonUnitaryAmp}). Here we study the same five resonances to test Eq.~(\ref{UnitaryTheta}). Unfortunately, unlike for the elastic scattering, in the inelastic case there are not enough results to make the world average. Therefore we use \v Svarc et al.~\cite{Sva14} results for pion photo-production. The agreement between our predictions and experimental values of $\theta$ can be seen in Table \ref{Table1}, where resonance parameters $M$, $\Gamma$, and $M_\mathrm{BW}$ are the estimates made by PDG \cite{PDG}, while $\beta$ and $\delta$s are calculated from them using Eqs.~(\ref{beta}) and (\ref{delta}) with \mbox{$W_{\pi N}=$ 1077 MeV}. Experimental $\theta$s and $x$s for $\pi N$ elastic are estimated from PDG data \cite{PDG}. Residue phases for $\gamma N \rightarrow \pi N$ are taken from Ref.~\cite{Sva14}, to which we added a phase of $\sqrt{k_{\gamma N}\, q_{\pi N}}$ calculated at the pole position due to different conventions ($k_{\gamma N}$ and $q_{\pi N}$ are c.m.~momenta).

\begin{table}[h!]
\caption{
Predictions of residue phase $\theta$ using the known resonance parameters and Eq.~(\ref{UnitaryTheta}).   \label{Table1}
 }
\begin{tabular}{lrrrrr} 
\hline\hline
	 								& ${\Delta(1232)}$	&  ${N(1520)}$ 		&  ${N(1675)}$   	&  ${N(1680)}$  	&  ${\Delta(1950)}$  \\ 
$J^\pi$								& $3/2^+$ 		& $3/2^-$ 			& $5/2^-$ 			& $5/2^+$ 		& $7/2^+$             \\
$L_{2I2J}^{\pi N}$ 						& $P_{33}$ 		& $D_{13}$ 		& $D_{15}$ 		& $F_{15}$ 		& $F_{37}$             \\
\hline
$M$/MeV 								& 1210$\pm$1  	& 1510$\pm$5	  	& 1660$\pm$5  	& 1675$\pm$10 	& 1880$\pm$10              \\ 
$\Gamma$/MeV						& 100$\pm$2  		& 110$\pm$10 		& 135$\pm$15	 	& 120$\pm$15		& 240$\pm$20             \\ 
$M_\mathrm{BW}$/MeV 					& 1232$\pm$2  	& 1515$\pm$5 		& 1675$\pm$5 		& 1685$\pm$5 		& 1930$\pm$20             \\
\hline
$\beta$/$^\circ$						& $-24$$\pm$2  		& $-5$$\pm$7 		& $-13$$\pm6$ 		& $-9$$\pm$10 	& $-23$$\pm$9              \\  
\hline
$x_{\pi N}$/\%							& 103$\pm$3 			& 61$\pm$5			& 39$\pm$6				& 69$\pm$10	& 43$\pm$5			\\	
$\delta_{\pi N}$/$^\circ$					& $-21$$\pm$0  		& $-7$$\pm$1 		& $-7$$\pm$1	 	& $-6$$\pm$1 		& $-8$$\pm$1           \\    
$\theta_{\pi N \pi N}$/$^\circ$				& $-44$$\pm$2 	 	& $-12$$\pm$7 		& $-19$$\pm$6 	& $-15$$\pm$10 	& $-31$$\pm$9             \\    
$\theta^{\mathrm{\, exp}}_{\pi N \pi N}$/$^\circ$	& $-46$$\pm$2	  	& $-10$$\pm$5 		& $-25$$\pm$6 	& $-10$$\pm$10 		& $-32$$\pm$8            \\    
\hline
$\delta_{\gamma N}$/$^\circ$						& $-10\pm$0  		& $-5$$\pm$1 		& $-5$$\pm$1 		& $-5$$\pm$1 		& $-7$$\pm$1           \\    
$\theta_{\gamma N \pi N}$/$^\circ$					& $-39$$\pm$2	  	& $-12$$\pm$7 	& $-19$$\pm$6 	& $-15$$\pm$10 	& $-30$$\pm$9           \\    
$\theta^{\mathrm{\, exp}}_{\gamma N \pi N}$/$^\circ$	& $-37$$\pm$2	  	& $11$$\pm$3 		& N/A 			& $-9$$\pm$3	 	& $-22$$\pm$3            \\      
 \hline\hline
\end{tabular}
\end{table}

\begin{figure*}[h!]
\begin{center}
\includegraphics[width=1\textwidth]{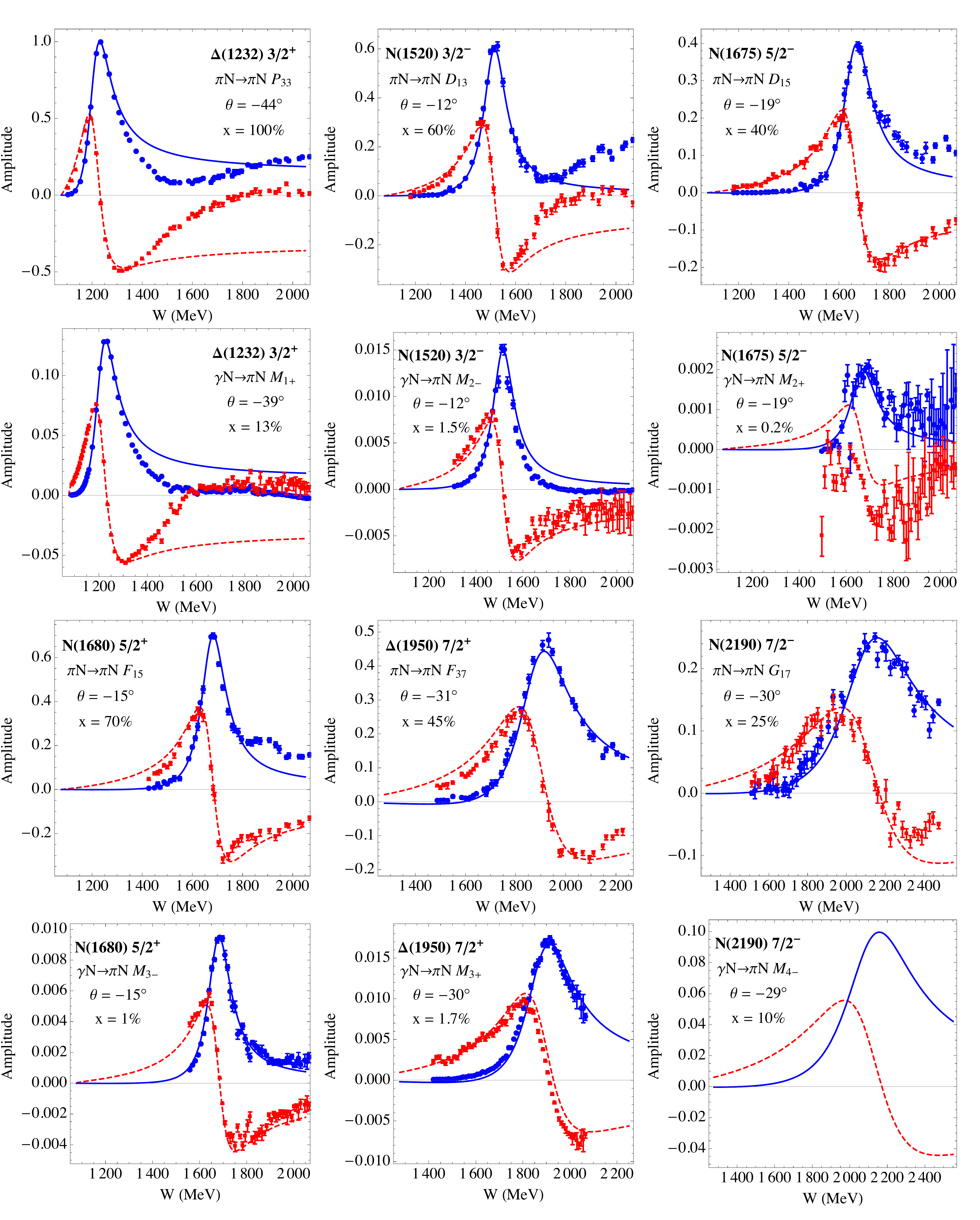}
\caption{ The real (dashed, red online) and imaginary (solid, blue online) parts of the resonant amplitude, built from Eq.~(\ref{UnitaryAmp}), are compared to the corresponding data points from SAID database \cite{SAID}. There was no photo-production data for N(2190). We did not fit the data in any way, apart from slightly adjusting the overall multiplicative constant $x$ within the error bars (compare with Table \ref{Table1}). The $\pi N$ elastic amplitudes and photo-production magnetic multipoles $M_{l\pm}$ are given in different normalizations. To make them comparable we multiply all $M_{l\pm}$ data with conversion function $10^{-3}/m_{\pi}\sqrt{2\,k_{\gamma N}\, q_{\pi N}}$.  \label{Fig2}}
\end{center}
\end{figure*}

The agreement between predicted and experimental $\theta$ values is very good in almost all cases. To get a better idea about this agreement, we construct real and imaginary parts of the amplitude using Eq.~(\ref{UnitaryAmp}) with parameters from \mbox{Table \ref{Table1}} (no fitting, $x$ slightly adjusted) and compare them to the $\pi N$ elastic and the photo-production single-energy data from SAID \cite{SAID} in Fig.~\ref{Fig2}. For the photo production we use the magnetic multipoles $M_{l\pm}$ multiplied with $10^{-3}/m_{\pi}\sqrt{2\,k_{\gamma N}\, q_{\pi N}}$.

In Fig.~\ref{Fig2} we included also N(2190) $7/2^-$ with averaged PDG parameter values: $M$ is 2100 MeV, $\Gamma$ is 439 MeV, and $M_\mathrm{BW}$ is 2171 MeV. This resonance is very broad and has a small branching fraction $x$ of only 25\%. The predicted residue phase of $-30^\circ$ is fully consistent with the $-30^\circ$ of Cutkosky et al.~\cite{Cutkosky}, $-32^\circ$ by Arndt et al.~\cite{Arndt}, and not too far from $-18^\circ$ by \v Svarc et al.~\cite{Sva14PDG}. Interestingly, it has the opposite sign from Sokhoyan et al.~\cite{Sok15} result of $30^\circ$. Making a reasonable estimate or average of these results presents quite a challenge for PDG; their estimate is $0^\circ\pm30^\circ$. That is another reason why plots might be better than tables.

\begin{figure}[h!]
\begin{center}
\includegraphics[width=0.5\textwidth]{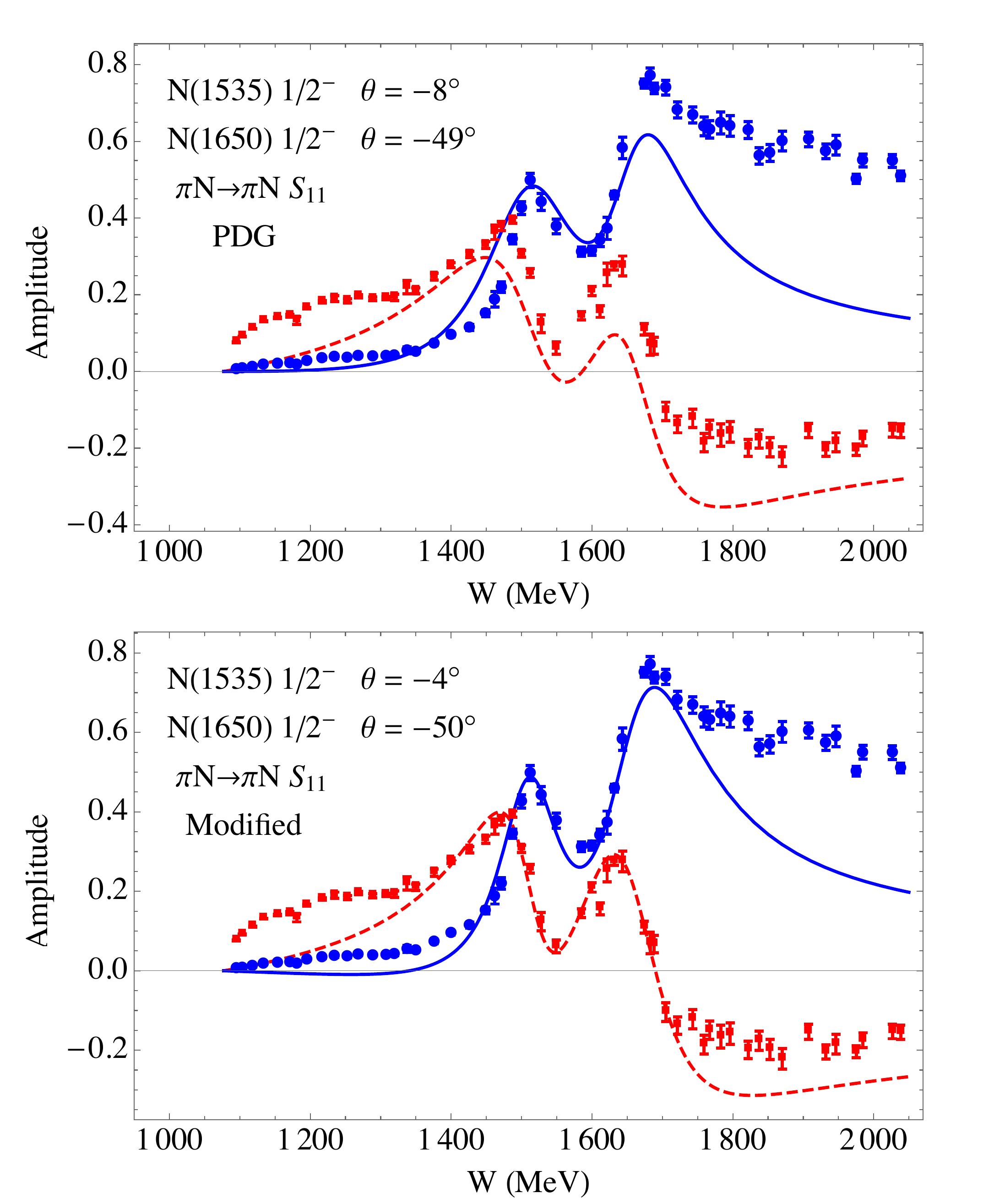}
\caption{  Combined $\pi N$ elastic amplitude for N(1535) and N(1650). Data is from SAID \cite{SAID}. In the upper figure we use the PDG parameters, while in the lower we modifiy some to better fit the data. Phases $\theta$ are determined numerically. \label{Fig3}}
\end{center}
\end{figure}

The proposed model is built for isolated resonances and as such it is bound to fail in describing overlapping resonances. Still, in Ref.~\cite{Cec16} it was shown that combining single resonance amplitudes $A_1$ and $A_2$ as \mbox{$A_1+A_2+2iA_1A_2$} produces a reasonable $\theta$ estimate for N(1535) and N(1650). In the upper part of Fig.~\ref{Fig3} we show how the combined amplitude roughly resembles the data.

To test how much $\theta$ predictions would change if there was a better fit to the data, we adjust some resonant parameters in the lower part of Fig.~\ref{Fig3}. It turns out, not too much. The parameters we use are given in Table \ref{Table2}.

 \begin{table}[h!]
\caption{Parameters we use for N(1535) and N(1650). Phase $\theta^{\mathrm{exp}}_{\pi N}$ is from PDG \cite{PDG}  (row PDG). \v Svarc et al.~\cite{Sva14PDG} residue phase is shown for comparison  (row Modified).\label{Table2} }
\begin{tabular}{lllllrr} 
\hline\hline
 & $x$ & $M$ & $\Gamma$ & $M_\mathrm{BW}$ & $\theta$ & $\theta^{\mathrm{exp}}_{\pi N}$\\
		  & (\%) & (MeV) &  (MeV) & (MeV) & ($^\circ$) & ($^\circ$)\\
		  	\hline
$^\mathrm{N(1535)1/2^-}_\mathrm{PDG}$			& 45$\pm$10 		& 1510$\pm$20  	&  170$\pm$80 	&  1535$\pm$25 	& $-8$	& $-15$$\pm$15 \\
$_\mathrm{Modified}$ 			&  42 	&  1510 	& 110 	&  1535 	& $-4$ 	& $-5$$\pm$6  \\
\hline
$^\mathrm{N(1650)1/2^-}_\mathrm{PDG}$			& 60$\pm$10		& 1655$\pm$15 	&  135$\pm$35  	&  1655$\pm^{15}_{10}$ 	& $-49$ 	& $-70$$\pm$20 \\
$_\mathrm{Modified}$		& 76 		&  1655  	&  160 	&  1680 	& $-50$ 	& $-47$$\pm$3 \\
\hline\hline
\end{tabular}
\end{table}

Before concluding, we would like to address a curious feature of a unitary and analytic coupled-channel multi-resonant analysis \cite{Batinic}. In Ref.~\cite{Cec08} it was shown that close to a resonance the trace of the multichannel matrix $A$ from \cite{Batinic} looked like a simple elastic single-resonance amplitude, with the peak position of its imaginary part (and zero of the real) consistent with the Breit-Wigner mass. That is rather intriguing since the peak positions of the amplitude's matrix elements in the presented model, or any other model have nothing to do with corresponding $M_\mathrm{BW}$. But, we are looking at the trace of the matrix, not at mere matrix elements. Simply calculating the trace of Eq.~(\ref{UnitaryAmp}) would not be too useful because the result is too general and we cannot learn much from it. That changes dramatically if we rewrite Eq.~(\ref{UnitaryAmp}) as 
\begin{equation}
A_{if} = X_{if}\,\sin(\rho+\beta)\,e^{i(\rho+\beta)},
\end{equation}
where real $X$ matrix absorbs everything else. When we impose the unitarity relation $A^\dag A=\mathrm{Im}\,A$, we get the relation $X^2=X$. Mathematically, this means that the trace of $X$ will be equal to its rank. The rank of a matrix is simply a number of its linearly independent rows (or columns). Consequently, the imaginary part of the trace of $A$ will be maximal (and the real part zero) when $\rho+\beta$ equals 90$^\circ$, which is exactly the definition of $M_\mathrm{BW}$.

\section{Conclusions}

We proposed a simple formula for multichannel scattering amplitude with parameters related to physical resonant properties. It can be used to predict the residue phase from other resonant parameters and describe the shape of experimentally determined amplitudes without any background contributions or fitting. In particular, we explained why $\Delta$(1232) has $\pi N$ elastic residue phase much lager (in magnitude) than most other prominent resonances: angle $\delta_{\pi N}$ is large because this resonance is the closest one to $\pi N$ threshold, angle $\beta$ is large because this resonance has large difference between $M$ and $M_\mathrm{BW}$, and $\theta$ is the sum of the two. Moreover, we predicted that the residue phase in the pion photo-production will be smaller because $\gamma N$ threshold is further away and angle $\delta_{\gamma N}$ is smaller. The proposed formula seems to make sense even for overlapping resonances, and residue phase extraction appears to be robust in respect to small adjustments of the resonance parameters. Having said that, we would not recommend this formula for extraction of resonant properties since it is a local (near-resonance) approximation which works well only for prominent resonances. Yet, it seems to be very useful in predicting and explaining typical behavior of resonant amplitudes and physical properties of resonances. After all, imposing unitarity to the formula did enable us to understand puzzling features (peaks and zeros coinciding with Breit-Wigner masses) of a much more advanced unitary model.

\section{Acknowledgements}

S.C.~would like to thank Lothar Tiator and Hedim Osmanovi\' c for invaluable discussions, comments, and suggestions.


\end{document}